%Paper: quant-ph/9503019
%From: E.J.Squires@durham.ac.uk
%Date: Mon, 27 Mar 1995 15:26:34 +0100 (BST)

\hsize=4.8 in
\hoffset=.75 in
\vsize=8 in
\voffset=.5 in
\rightline {DTP/95/13}

 \vskip 1cm
 \font\large=cmbx12 at 20pt
 \centerline{\large{Gravity, Energy Conservation}}
\vskip .4cm  \centerline{\large {and}}
\vskip .4cm
\centerline{\large { Parameter Values in
Collapse Models}}
\bigskip
\bigskip
\centerline{Philip Pearle}
\centerline{Hamilton College}
\centerline{Clinton, New York 13323, U.S.}
\medskip
\centerline{and}
\medskip
\centerline{Euan Squires}
\centerline{Department of Mathematical Sciences}
\centerline{University of Durham}
\centerline{Durham City, DH1 3LE, U. K.}
\bigskip
\centerline {March, 1995}
\vskip 1cm
\centerline{\bf Abstract}

	We interpret the probability rule of the CSL collapse theory to mean that
the scalar field
which causes collapse is the gravitational curvature scalar with two
sources,
the expectation value of the mass density (smeared over the GRW scale $a$)
and a white noise
fluctuating source.  We examine two models of the fluctuating source,
monopole fluctuations and
dipole fluctuations, and show that these correspond to two well known CSL
models.  We
relate the two GRW parameters of CSL to
fundamental constants, and we explain the energy increase of particles due
to collapse as
arising from the loss of vacuum gravitational energy. It is shown how
a problem with semi-classical gravity may be cured when it is combined with
a CSL collapse model.

\bigskip

\font\bfg=cmbx12
\noindent {\bfg {1. Introduction}}

\bigskip

	In collapse
models the Schr\"odinger equation is modified
so that it describes collapse of the wavefunction as a dynamical process.
This modification is introduced for the best possible reason,
namely, that the current theory is unable to account properly for
experiments.
In this respect the inadequacy of standard quantum theory is subtle:
the perfect agreement with all present experiments is only obtained through
a
crude ``instantaneous collapse'' prescription, which has no explanation
within the theory.$^1$

In the Continuous Spontaneous Localisation (CSL) theory $^{2-4}$ (see
section 2),
collapse is caused by interaction of the quantum system with a classical
scalar field,
$w({\bf x}, t)$. The theory probably gives the best
description of collapse available at the present time
but, inevitably, although it solves some problems it
produces others. In particular, it has been criticised for
three reasons. The first is that the collapse narrows wavefunctions,
thereby
producing an increase of energy$^5$ (see section 3),
which raises
the question as to whether there
is a violation of energy conservation, or whether this energy has some
as yet unspecified source.

The second criticism is that the nature of
the important physical field $w$ is not specified and, in particular,
it is not associated with anything else in physics. Related to this
is the third criticism, which is that the two parameters which
specify the model are ad hoc. These two parameters,
which were originally introduced in the seminal work of
Ghirardi, Rimini and Weber,$^6$ are a distance scale,
$a\simeq 10^{-5}$ cm, characterising the
distance beyond which the collapse becomes effective,
and a time scale, $\lambda ^{-1} \simeq 10^{16}$ sec,
giving the rate of collapse for, say, a proton.
If collapse is a fundamental physical process
related to other fundamental processes, it might be expected that
the parameters can be written in terms of other physical constants.

	In this note we shall address all three of these problems.

	  There are two equations which characterize CSL.
The first equation is a modified Schr\" odinger equation, which
expresses the influence of an arbitrary field $w({\bf x},t)$ on the quantum
system.
But it is the second equation which stimulates this paper.  This equation
is a probability rule which gives the
probability that nature actually chooses a particular $w({\bf x},t)$.
However, this
probability rule can be interpreted as expressing the influence of the
quantum
system on the field $w({\bf x},t)$. It can be shown$^{3}$ (section 2)
to be completely equivalent to

	$$w({\bf x},t)=w_{0}({\bf x},t)+<A({\bf x},t)>\eqno(1.1)$$

\noindent where $<A({\bf x},t)>$ is the quantum
expectation value of the mass operator smeared over the distance $a$ (see Eq.
(2.2)), and
$w_{0}({\bf x},t)$ is a gaussian randomly fluctuating field with
zero drift, temporally white noise in character and with a particular
spatial correlation function.

 	In this paper we shall take Eq. (1.1) seriously, using it as a guide
to understanding how the collapse formalism fits into the rest of physics.
 First, it tells us that
$w({\bf x},t)$'s average value is $<A>$, a mass density.  We therefore are
led to write
$w({\bf x},t)$ as

	  $$w({\bf x},t)\equiv{1\over4\pi G}\nabla ^{2}\phi({\bf x},t)\eqno(1.2)$$

\noindent We note that if we interpret the $\phi$ defined in (1.2) as the
actual
gravitational potential (so $w$ is the Newtonian
limit of ${1\over 2}$ the spacetime curvature scalar),
we are doing two unconventional things.

	First, we are using semi-classical gravity$^{7}$ (because
it is the expectation value of $A$ that
is the source of $\phi$).

	 Moreover, because of the nature of $A$, we are
led to the strange notion that a point particle has an
effect on the gravitational potential as if its mass were smeared over the
GRW scale $a$. We do not
believe there is any experimental evidence which conflicts with such a
possibility,
so we shall entertain it.  Indeed,
we shall show how (see section 6) a collapse model combined with such a
smearing
can eliminate a possible source of inconsistency in semi-classical
gravity$^{2}$
by ensuring that a nonlocalized state collapses to a localized state
before the gravitational field of the nonlocalized state
can be measured (essentially, the smearing weakens
the gravitational field so that its measurement
is prolonged beyond the collapse time).

	With this interpretation of $w$, the presence of $w_{0}$ in Eq. (1.1)
informs us that the gravitational potential (and associated curvature) also
fluctuate.
We shall investigate two different naive
 classical models of the source of these fluctuations, a monopole model
(section 4)
 and a dipole model (section 5).

 	In the monopole model
we assume that the source consists of particles of mass $\mu$ which appear
at random
times and positions: each persists for a short fixed time interval $\cal T$
in a
fixed volume ${\cal L}^3$ which we refer to as a ``cell."  We shall also
assume that there is a fixed background of negative mass so that, in each
cell,
the fluctuations appear as positive and negative masses which average to
zero. (This is vaguely
like a classical model of virtual quantum fluctuations, with the negative
background
being a renormalization subtraction.)  We shall eventually take $\mu$ to be
the planck mass
($\mu\equiv (\hbar c/G)^{1\over 2}\approx 2.2\times10^{-5}$ gm.) and $\cal
T$ to be the planck time
(so that ${\cal T}=\hbar/  \mu c^{2}$ as expected for a quantum
fluctuation).

	 In the dipole model we assume that masses $\mu$ and $-\mu$ appear
at random but simultaneously in pairs which occupy adjacent cells with
random orientation.

	Having produced some assumptions which make a well defined break
with normal physics, we proceed to discuss the consequences using normal
(classical) physics.
We calculate the correlation function of $w$.  We find that
the correlation function for the monopole model is $\sim \delta ({\bf
x}-{\bf x}')
\delta (t-t')$, and the correlation function for the dipole model is
$\sim {\bf \nabla} \cdot{\bf \nabla}'\delta ({\bf x}-{\bf x}')\delta
(t-t')$.  These
are respectively identical in form to the
correlation functions obtained in the original CSL model$^{2-4}$ (hereafter
called
GRWP), and in the model
proposed by Diosi$^{8}$ and corrected by Ghirardi, Grassi and Rimini$^{9}$
(hereafter called DGGR).
Upon equating the correlation functions of the monopole model and GRWP
we obtain one equation relating the GRW parameters and
the parameters of the monopole model, and likewise for the dipole model and
DGGR.

	The fluctuating monopoles
or dipoles exert a random gravitational force
on a particle,
causing it to undergo undamped Brownian motion, with the result that, on
average,
its kinetic energy increases
linearly with time.$^{10}$
An average linear increase
of kinetic energy for each particle is precisely the behavior CSL predicts
to occur during
collapse. Thus, although we do not have a classical picture of the complete
collapse, we have
a classical picture for this aspect of it.
Upon equating the classical and CSL expressions for the energy increase,
we obtain a second equation relating the GRW parameters and
the parameters of the monopole model, and likewise for the dipole model.
We shall examine the consequences of these two equations for the GRW
parameters, for each model.

	Thus we have a more---or---less plausible argument that the
collapse mechanism is gravitationally related, that the energy for collapse
comes from the vacuum, and that the GRW parameters can be related
to fundamental constants.

	We expect that the reader will, as we do, take the argument given here
with a grain of salt. For example, it is very likely that a better
justified
picture of curvature fluctuations can be found.
But, at the least, our arguments indicate
 that the aforementioned criticisms of CSL
are not insurmountable and that their solutions may point the way
toward a more complete theory. At most, it is not inconceivable that the
scale of fluctuations
necesary for collapse as well as other elements given
here may survive in such a theory.  Indeed, ideas of earlier
authors appear here.  Karolyhazy$^{11}$ proposed that metric fluctuations
play a role in collapse.
Penrose$^{12}$ has for many years argued that gravity and
collapse are linked.  Diosi$^{8}$ was the first to write an expression for
a GRW
parameter in terms of fundamental constants in a CSL-gravitational model:
he had hoped to do without
the mass smearing whose necessity was
pointed out by Ghirardi, Grassi and Rimini.$^{9}$ The
possibility that collapse might cure a crucial problem of semi-classical
gravity,
which was suggested
in the first paper on CSL,$^{2}$ is further developed here.

\bigskip

\noindent {\bfg {2. CSL}}

\bigskip

	We shall begin with the solution of the general
nonrelativistic CSL \break Schr\" odinger
evolution equation for the statevector in the interaction picture, under
the
influence of an arbitrary scalar field w({\bf x},t):$^{13}$

$$|\psi,T>_{w}={\bf T}e^{-\int_{0}^{T}dt\int\int d{\bf x}d{\bf x}'[w({\bf
x},t)-A({\bf x},t)]
G^{-1}({\bf x}-{\bf x}')[w({\bf x}',t)-A({\bf x}',t)]}|\psi,0>\eqno(2.1)$$

\noindent (${\bf T}$ is the time ordering operator). In Eq. (2.1), for
fixed t,
$A({\bf x},t)$ is an {\bf x}-parameter labelled family
of commuting (interaction picture) operators toward
whose joint eigenstates the collapse tends during the interval $(t,\
t+dt)$.
In light of recent discussions of experiments,$^{14,15}$
we take $A$ to be the mass density operator (smeared over $a$):

$$A({\bf x},t)\equiv e^{iHt}e^{{1\over 2}a^{2}\nabla ^{2}}
M({\bf x})
e^{-iHt}\eqno(2.2)$$

\noindent where the mass density operator is

$$M({\bf x})\equiv\sum_jm_{j}\xi^{\dagger}_j({\bf x})\xi_j({\bf x})
\eqno(2.2a)$$

\noindent ($\xi_j({\bf x})$ is the annihilation operator for a particle of
mass $m_{j}$
at the point {\bf x}) and the smearing is described by

$$e^{{1\over 2}a^{2}\nabla ^{2}}M({\bf x})={1\over (2\pi a^{2})^{3\over
2}}\int d{\bf z}
e^{-{1\over 2a^{2}}({\bf x}-{\bf z})^{2}}M({\bf z})\eqno(2.2b)$$

\noindent It is crucial that the mass density operator $M({\bf z})$
be smeared over the scale $a$, as indicated in Eq. (2.2b). Without such a
smearing
the energy excitation of particles undergoing collapse would be beyond
experimental constraints.

	In Eq. (2.1), $G^{-1}({\bf x}-{\bf x}')$ is a real positive definite
function of
$|{\bf x}-{\bf x}'|$.  In later sections we shall be concerned with two
particular examples:

$$\eqalign{\hbox{GRWP:\qquad  }G^{-1}({\bf x}-{\bf x}')
& =\gamma\delta ({\bf x}-{\bf x}')\cr
\hbox{DGGR:\qquad  }G^{-1}({\bf x}-{\bf x}')
& =\gamma '{1\over |{\bf x}-{\bf x}'|}\cr}\eqno(2.3a,b)$$

\noindent where $\gamma$ or $\gamma '$ is a constant.
For later purposes it is useful to define here the
inverse of $G^{-1}({\bf x}-{\bf x}')$, namely $G({\bf x}-{\bf x}')$:

$$\int d{\bf z}G({\bf x}-{\bf z})G^{-1}({\bf z}-{\bf x}')=
\delta ({\bf x}-{\bf x}')\eqno(2.4)$$

\noindent It follows from Eq. (2.4) that the (positive real) fourier
transforms of
$G$ and $G^{-1}$ are reciprocals. For the two cases,

	$$\eqalign{\hbox{GRWP:\qquad  }G({\bf x}-{\bf x}')
	&={1\over \gamma}\delta ({\bf x}-{\bf x}')\cr
\hbox{DGGR:\qquad  }G({\bf x}-{\bf x}')&=-{1\over 4\pi\gamma '}
\nabla^{2}\delta ({\bf x}-{\bf x}')\cr}\eqno(2.5a,b)$$

	The probability rule of CSL, giving the probability
that nature chooses a particular $w({\bf x},t)$ for $0\leq t \leq T$, is

$$\hbox {Prob}\{w({\bf x},t)\}=Dw\thinspace_{w}
\negthinspace\negthinspace <\psi,T|\psi,T>_{w}\eqno(2.6)$$

\noindent where Dw is the functional integral element

$$Dw\equiv C\prod_{{\rm all\ }{\bf x}:\ t=0}^{t=T}dw({\bf
x},t)\eqno(2.6a)$$

\noindent and C, proportional to $(\det G)^{-{1\over 2}}$, makes the
integrated probability =1.

	Starting with the  probability rule (2.6) we may take the following steps,
justified
because the terms we drop are of negligible order in dt (i.e., they make no
contribution to the expectation values of functionals of $w$):

$$\eqalign{&\hbox{Prob}\{w({\bf x},t)\}=\cr
 &\qquad Dw\prod_{t=0}^{T}{\thinspace_{w}
\negthinspace\negthinspace<\psi,t|
e^{-2dt\int d{\bf x}d{\bf x}'[w({\bf x},t)-A({\bf x},t)]
G^{-1}({\bf x}-{\bf x}')[w({\bf x}',t)-A({\bf x}',t)]}|\psi,t>_{w}
\over\thinspace_{w}
\negthinspace\negthinspace<\psi,t|\psi,t>_{w}}\cr
&\qquad\qquad=Dw\prod_{t=0}^{T}e^{-2dt\int d{\bf x}d{\bf x}'w({\bf x},t)
G^{-1}({\bf x}-{\bf x}')w({\bf x}',t)}\cr
&\qquad\qquad\qquad\qquad\cdot\Bigl[1+4dt\int d{\bf x}d{\bf x}'G^{-1}({\bf
x}-{\bf x}')
{\thinspace_{w}
\negthinspace\negthinspace<\psi,t|A({\bf
x}',t)|\psi,t>_{w}\over\thinspace_{w}
\negthinspace\negthinspace<\psi,t|\psi,t>_{w}}\Bigr]\cr
&\qquad\qquad=Dw
 e^{-2\int _{0}^{T} dt\int\int d{\bf x}d{\bf x}'[w({\bf x},t)-<A({\bf
x},t)>]
G^{-1}({\bf x}-{\bf x}')[w({\bf x}',t)-<A({\bf x}',t)>]}}\eqno (2.7a,b,c)$$
$$<A({\bf x},t)>\equiv {\thinspace_{w}
\negthinspace\negthinspace<\psi,t|A({\bf x},t)|\psi,t>_{w}\over
\thinspace_{w}
\negthinspace\negthinspace <\psi,t|\psi,t>_{w}}\eqno(2.7d)$$

\noindent  It follows from Eq. (2.7c) that the probability rule is
completely equivalent
to the expression (1.1) for $w({\bf x},t)$,
where $<A({\bf x},t)>$ is given by (2.7d), and $w_{0}({\bf x},t)$ is a zero
drift gaussian
process characterized by the correlation function

$$<w_{0}({\bf x},t)w_{0}({\bf x}',t)>
={1\over 4}G({\bf x}-{\bf x}')\delta (t-t') \eqno(2.8)$$

	The density matrix may be found from Eqs. (2.1), (2.6):

$$\rho (T) =\int Dw\thinspace_{w}
\negthinspace\negthinspace <\psi,T|\psi,T>_{w}{|\psi,T>_{w}\thinspace_{w}
\negthinspace\negthinspace <\psi,T|\over \thinspace_{w}
\negthinspace\negthinspace <\psi,T|\psi,T>_{w}}
 =\int Dw |\psi,t>_{w}\thinspace_{w}
\negthinspace\negthinspace <\psi,t|$$
$$={\bf T}e^{-{1\over 2}\int_{0}^{T}dt\int\int
d{\bf x}d{\bf x}'[A({\bf x},t)\otimes 1-1\otimes A({\bf x},t)]
G^{-1}({\bf x}-{\bf x}')[A({\bf x}',t)\otimes 1-1\otimes A({\bf
x'},t)]}\rho (0)\eqno(2.9)$$

\noindent (the notation is $(A\otimes B)C\equiv ACB$, and the ${\bf T}$
operator is
time-reverse ordering for operators to the right of $\otimes$).

	Before concluding this section, we wish to make one further point.
It may have occurred to the reader that there is freedom to transform $w$,
with a concommitant transformation of $A$ and $G^{-1}$ so that
the exponent in Eq. (2.1) is left unchanged.
In particular, one may think of defining
$w({\bf x},t)\equiv \exp[a^{2}\nabla ^{2}/2]w'({\bf x},t)$, with
the result that, in (2.1), $w({\bf x},t)$ is replaced by
$w'({\bf x},t)$, $A({\bf x}',t)$ is replaced by $M({\bf x}',t)$, and
$G^{-1}({\bf x}-{\bf x}')$ is replaced by

	$$\eqalign{G'^{-1}({\bf x}-{\bf x}')&\equiv e^{a^{2}\nabla
^{2}}G^{-1}({\bf x}-{\bf x}')\cr
	&={1\over (4\pi a^{2})^{3\over 2}}\int d{\bf z}
e^{-{1\over 4a^{2}}({\bf x}-{\bf x}'-{\bf z})^{2}}G^{-1}({\bf z})\cr
&={1\over (2\pi)^{3\over 2}}\int d{\bf k}e^{i{\bf k}\cdot({\bf x}-{\bf
x}')}
e^{-a^{2}k^{2}}\tilde G^{-1}({\bf k})\cr}\eqno(2.10a,b,c)$$

\noindent This would result in Eq. (1.1) being replaced by $w'=w'_{0}+M$,
where $M$ is the mass
density.  It would appear that the smeared mass density $A$ to which we
attributed significance
in section 1 would be unnecessary.  What's wrong with this?  It is that
this
transformation cannot be allowed because $w'_{0}$
is undefined:  $G'$, the correlation function of $w'_{0}$, is equal to the
fourier transform  of
$\exp[a^{2}k^{2}]\tilde G({\bf k})$ which does not exist.

	However, there are transformations which are allowed. For example, $w\sim
\nabla^{2}\phi$
can be replaced by $\phi$, $A({\bf x},t)$ by
$\sim\nabla^{-2}A({\bf x},t)=-\int d{\bf z}A({\bf z},t)/4\pi |{\bf x}-{\bf
z}|$,
and $G^{-1}({\bf x}-{\bf x}')$ by $\sim \nabla ^{2}\nabla '^{2}G^{-1}({\bf
x}-{\bf x}')$. Since
the models obtained under such allowed transformations are equivalent
we could, for example, discuss fluctuations of $\phi$ instead of
fluctuations
of $w=\nabla ^{2}\phi$, but scalar field seems
less fundamental than curvature and the CSL expressions are simplest for
$w\sim\nabla ^{2}\phi$.

\bigskip

\noindent {\bfg {3. Energy Production}}

\bigskip

	We shall calculate the energy production rate which accompanies collapse.
By taking the time derivative of Eq. (2.9), we see that
$\rho$ satisfies the differential equation

$${d\rho (t)\over dt}=-{1\over 2}\int\int d{\bf x}d{\bf x}'G^{-1}({\bf
x}-{\bf x}')
[A({\bf x},t),[ A({\bf x}',t),\rho (t)]]\eqno(3.1)$$

\noindent In the position basis (writing $|x>$ for the position eigenvector
of all particles),
Eq. (3.1) is

$$\eqalign{&{d<x|\rho (t)|x'>\over dt}=-{1\over 2}e^{i{\cal H}t}
\sum_i\sum_j\cr &\quad[\Phi({\bf x}_{i}-{\bf x}_{j})
+\Phi({\bf x}'_{i}-{\bf x}'_{j})-2\Phi({\bf x}_{i}-{\bf x}'_{j})]
e^{-i{\cal H}t}<x|\rho (t)|x'>\cr}\eqno (3.2)$$

\noindent where

$$\Phi({\bf x}_{i}-{\bf x}_{j})\equiv{m_{i}m_{j}\over (4\pi a^{2})^{3\over
2}}\int
d{\bf z}G^{-1}({\bf z})e^{-{1\over 4a^{2}}[{\bf z}-({\bf x}_{i}-{\bf
x}_{j})]^{2}}\eqno(3.3)$$

\noindent so that for the two cases of interest (using Eqs.(2.3)),

$$\eqalign{\hbox{GRWP:\qquad}\Phi({\bf x}_{i}-{\bf x}_{j})
	&=\biggl\{{\gamma m_{i}m_{j} \over(4\pi a^{2})^{3\over 2}}\biggr\}
	e^{-{1\over 4a^{2}}[{\bf x}_{i}-{\bf x}_{j}]^{2}}\cr
\hbox{DGGR:\qquad}\Phi({\bf x}_{i}-{\bf x}_{j})
	&=\biggl\{{\gamma 'm_{i}m_{j} \over a\pi^{1\over 2}}\biggr\}{1\over |{\bf
x}_{i}-{\bf x}_{j}|}
	\int_{0}^{|{\bf x}_{i}-{\bf x}_{j}|}
	dze^{-{1\over 4a^{2}}z^{2}}\cr}\eqno(3.3a,b)$$

\noindent We note that the bracketed expressions in Eqs. (3.3a,b) have the
dimension of 1/time.

	According to Eq. (3.2), the collapse rate for a ``pointer,"
composed of $N$ particles of mass $m$ with
${\cal M}\equiv Nm$ and mass density D,
in a state which is a superposition of the pointer in two locations
separated by a large distance
(much greater than $a$ or the pointer size $L$) is
 $\sim\sum_{i,j}\Phi({\bf x}_{i}-{\bf x}_{j})$.  From (3.3) we obtain

$$\eqalign{&\hbox{GRWP:\qquad}\hbox{Collapse Rate}\sim{\gamma{\cal
M}^{2}\over a^{3}}
\ (L<a),\qquad\sim\gamma {\cal M}D\ (L>a)\cr
&\hbox{DGGR:\qquad}\hbox{Collapse Rate}\sim{\gamma '{\cal M}^{2}\over a}
\ (L<a),\qquad\sim{\gamma '{\cal M}^{2}\over L}\ (L>a)\cr}\eqno(3.4a,b)$$

\noindent  These results will be useful in section 6.

	To find the average rate of energy increase, we
multiply Eq. (3.2) by the Hamiltonian ${\cal H}$,
and take the trace:

$${d\over dt}{\rm Tr}[H\rho (t)]=\sum_{j}{3\hbar^{2}\over
4m_{j}a^{2}}\lambda_{j}\eqno(3.5)$$
$$ \lambda_{j}\equiv {2m_{j}^{2}a^{2}\over 3(2\pi)^{3\over 2}}\int
d{\bf k}k^{2}e^{-a^{2}k^{2}}{\tilde G}^{-1}(k^{2})\eqno (3.5a)$$

\noindent (${\tilde G}^{-1}(k^{2})$ is the fourier transform of $G^{-1}$).
$\lambda_{j}$
is the GRW collapse rate for a single particle of mass $m_{j}$. For our two
cases of
special interest (2.3a,b), ${\tilde G}^{-1}= \gamma/(2\pi)^{3\over 2}$
and $(2/\pi)^{1\over 2}\gamma/k^{2}$
respectively, so

$$\eqalign{\hbox{GRWP:\qquad}\lambda_{j}
	&={m_{j}^2\gamma\over (4\pi)^{3\over 2} a^{3}}\cr
\hbox{DGGR:\qquad}\lambda_{j}
	&={m_{j}^2\gamma '\over 3\pi^{1\over 2}a}\cr}\eqno(3.6a,b)$$

 	It is remarkable that the average rate of energy increase (3.5)
is independent of the quantum state of the particles (as well
as their interaction potential). This makes possible the classical
modelling of
this energy increase presented in the next two sections.

\bigskip

\noindent {\bfg {4. Monopole Model}}

\bigskip

	For simplicity, imagine space partitioned into cubical cells of
edge length $\cal L$, and time divided into intervals of duration ${\cal
T}$.
Let $u_{\alpha, \beta}({\bf z},t)=1$ if a monopole
is in the $\alpha$th cell during the $\beta$th interval, for ${\bf z}$ in
the cell and $t$
in the interval, and $u_{\alpha, \beta}({\bf z},t)=0$ otherwise. Suppose
the monopole mass $\mu$ uniformly fills the cell when it appears.

	Denote by $\cal P$ the probability that a monopole
appears in any cell during
the $\beta$th interval. In addition, let there be a constant mass density
$-\mu {\cal P}/{\cal L}^{3}$ throughout
space, so that on average there is zero mass in each cell.  The potential
at ${\bf x}$ when $t$ lies in the $\beta$th interval, due
to all the fluctuating monopoles and the constant mass density, is
therefore

$$\phi({\bf x},t)=-G\mu\sum_{\alpha}{1\over{\cal L}^{3}}\int d{\bf z}
{[u_{\alpha,\beta}({\bf z},t)-{\cal P}]\over |{\bf x}-{\bf z}|}\eqno
(4.1)$$

\noindent  We note that $<\phi ({\bf x},t)>=0$, since
$<u_{\alpha,\beta}({\bf z},t)>=\cal P$.

	The correlation function of the potential is

	$$\eqalign{&<\phi ({\bf x},t)\phi ({\bf x}',t')>=\cr
&\quad\delta_{\beta \beta '}\Theta_{\beta}(t)\Theta_{\beta}(t')
(G\mu)^{2}{{\cal P}(1-{\cal P})\over({\cal L})^{6}}\sum_{\alpha}\int \int
d{\bf z}d{\bf z}'
{\Theta_{\alpha}({\bf z})\Theta_{\alpha}({\bf z}')\over|{\bf x}-{\bf
z}||{\bf x'}-{\bf z'}|}\cr
&\qquad\qquad\qquad\qquad\approx \delta (t-t')
(G\mu)^{2}\tilde{\cal P}\int d{\bf z}
{1\over|{\bf x}-{\bf z}||{\bf x'}-{\bf z}|}\cr}\eqno(4.2a,b)$$

\noindent with dependence only upon the combination of
model parameters $\tilde{\cal P}\equiv {\cal P}{\cal T}/{\cal L}^{3}$.
We have also introduced the characteristic function $\Theta_{\alpha}({\bf
z})$
of the $\alpha$th cell, which equals 1 if ${\bf z}$ lies in the cell, and 0
otherwise.
Similarly, $\Theta_{\beta}(t)$ is the characteristic function for the
$\beta$th time interval.  In
Eq. (4.2b) we have made some approximations:  ${\cal P}<<1$,
$\delta_{\beta \beta '}\Theta_{\beta}(t)\Theta_{\beta}(t')
\approx {\cal T}\delta (t-t')$ and
$\Theta_{\alpha}({\bf z})\Theta_{\alpha}({\bf z}')\approx
({\cal L})^{3}\delta ({\bf z}-{\bf z}')$.

	Eq. (4.2b) is all we shall need to calculate all quantities of interest
(actually, the integral
on the right hand side of (4.2b) must be cut off
at large $|{\bf z}|$ to exist, but the quantities we shall
calculate need no cutoff).  The correlation function of
$w_{0}({\bf x},t)=(4\pi G)^{-1}\nabla^{2}\phi({\bf x},t)$ is easily
obtained:

	$$<w_{0}({\bf x},t)w_{0}({\bf x}',t')>= \mu^{2}\tilde{\cal P}\delta (t-t')
	\delta ({\bf x}-{\bf x}') \eqno (4.3)$$

\noindent We see from Eqs. (2.5a) and (2.8) that these monopole
fluctuations give us the GRWP
correlation function, with $\gamma =1/4\mu^{2}\tilde{\cal P}$.
We thus obtain from Eq. (3.6a) an
expression for the collapse rate for a particle of mass $m$:

	$$\lambda_{m}={m^{2}\over 32\pi^{3\over 2}\mu^{2}a^{3}\tilde{\cal P}}\eqno
(4.4)$$

	We now turn to calculate the rate of energy increase of a particle of mass
m due to
the force it feels from the fluctuating potential.  As explained in section
1,
the particle causes its own gravitational potential as if its mass were
smeared over
the scale $a$ so, for consistency, when we calculate the force on a
particle at the origin, we
treat it as a rigid mass distribution
of density $m(2\pi a^2)^{-{3\over 2}}\exp -x^{2}/2a^{2}$. The correlation
function of the
force follows from Eq. (4.2b):

$$<F^{i}(t)F^{j}(t')>=
\delta (t-t'){(G\mu m)^{2}\tilde{\cal P}\over(2\pi a^2)^{3}}
\int d{\bf x}d{\bf x}'d{\bf z}
e^{-{1\over 2a^{2}}(x^{2}+x'^{2})}
\partial_{i}\partial '_{j}{1\over|{\bf x}-{\bf z}||{\bf x'}-{\bf z}|}
\eqno (4.5a)$$

\noindent Using the symmetry of the integrand, $\partial_{i}\partial '_{j}$
may be replaced
by $-(1/3)\delta_{ij}\nabla_{z}^{2}$ which, acting on $|{\bf x}-{\bf z}|$,
gives a delta function.
The remaining integral is straightforward, and the result is

$$<F^{i}(t)F^{j}(t')>=
\delta (t-t')\delta_{ij}{4\pi^{1\over 2}(G\mu m)^{2}\tilde{\cal P}\over
3a}\eqno (4.5b)$$

	The correlation function in Eq. (4.5b) is that of white noise.
Therefore we can write $F_{i}(t)=KdB_{i}(t)/dt$, where
$K^{2}$ is the constant factor in (4.5b) and $B_{i}(t)$ is Brownian motion,

$<B_{i}(t)B_{j}(t)>=\delta_{ij}t$.   By Newton's second law, the momentum
is $KB_{i}(t)$,
so the energy is $E=K^{2}{\bf B}(t)\cdot{\bf B}(t)/2m$.  We thus obtain

$${d<E>\over dt}={2\pi^{1\over 2}m(G\mu)^{2}\tilde{\cal P}\over a}\eqno
(4.6)$$

	The effect of the
fluctuating field on the particle, according to both our classical
calculation (4.6)
and the CSL calculation (3.5), is a linear rate of increase of energy.
Equating
the two provides a second relationship involving $\lambda_{m}$ and $a$:

$${3\hbar^{2}\over 4ma^{2}}\lambda_{m}=
{2\pi^{1\over 2}m(G\mu)^{2}\tilde{\cal P}\over a}\eqno (4.7)$$

	Eqs. (4.4) and (4.7) may be solved for $\lambda_{m}$ and $a$:

	$$\eqalign{a&=\biggl({3\over \pi^{2}}\biggr)^{1\over 4}{1\over 4}
	\biggl({c\hbar\over G\mu^{2}}\biggr)^{1\over 2}
	\biggl({1\over\tilde{\cal P}c}\biggr)^{1\over 2}\cr
	\lambda_{m}&
	={1\over 2(3\pi)^{1\over 2}}{G m^{2}\over a\hbar}\cr}\eqno(4.8a,b)$$

\noindent The fact that $a$ is independent of $m$ (as required by the CSL
models proposed so far), and that $\lambda_{m}$ is proportional to $m^{2}$
(as required
by (3.6a)) may be regarded as modest successes of the model. It is also
interesting that $\lambda_{m}a$ is independent of the parameters of the
model.

	If we use the GRW value for $a$ ($10^{-5}$ cm) in (4.8b), then we find
$\lambda_{m}\simeq 10^{-24}$ sec$^{-1}$ for a nucleon.  While this is eight
orders of magnitude
smaller than the value given by GRW, it is not completely unreasonable.
Indeed,
the expression (4.8b) for $\lambda$ for a proton and the collapse rate for
objects
of size smaller than $a$ is the same as in DGGR (up to a numerical factor).

The collapse time for a cube .01 cm on a side in a superposition of states
with separation
larger than .01 cm is longer than in GRWP or DGGR, but still a respectable
$10^{-5}$ sec.

	We have no good argument for choosing $\tilde{\cal P}$
and $\mu$, and so determining $a$.  Nonetheless, it is
interesting to see what is implied if we use the planck mass for $\mu$ (so
we may set
$c\hbar/G\mu^{2}=1$ in (4.8a)), the planck time for $\cal T$,
and the GRW value for $a$. Then, by (4.8a),
${\cal P}/{\cal L}^{3}\approx (4a)^{-2}(\hbar/\mu c)^{-1}\approx 4\times
10^{41}$ cm$^{-3}$.  It
so happens that $1/(\hbar/M c)^{3}\approx 10^{41}$ cm$^{-3}$, where $M$ is
the nucleon mass.
Indeed, if we use ${\cal P}/{\cal L}^{3}=1/(\hbar/M c)^{3}$ in Eqs. (4.8)
we obtain

$$\eqalign{a&=({3\over \pi^{2}})^{1\over 4}
{\hbar\over 4Mc}\sqrt{\mu \over M}\approx 1.4\times 10^{-5}{\rm \ cm}\cr
\lambda_{m}&\approx 2\times10^{-24}
\hbox{\ sec$^{-1}$ for the nucleon}\cr}\eqno(4.9a,b)$$

	Thus we obtain the GRW value for $a$ provided the frequency of
appearance of the monopole ``planckons"
is such that on average there is always one present per proton volume
(taking the
compton radius of the proton to characterize its size).
This is a very suggestive number even though we have no obvious theory for
it.
It suggests that the existence of a particle of mass $m$ may cause
planckon fluctuations in a region of space around it with probability/vol
equal to
$1/(\hbar/mc)^{3}$. Then, in ordinary matter, the probability/vol is
dominated by
the planckons due to the presence of the nucleons, since the
probability/vol due to the
presence of electrons is $10^{-10}$
times smaller. (This smaller planckon production rate
would, of course, be obtained in a purely electron plasma).
This would make $a$ dependent upon the milieu in which particles find
themselves,
and would represent a variant of standard CSL, where it has been assumed up
to now that $a$ is universal.

\bigskip

\noindent {\bfg {5. Dipole Model}}

\bigskip

	We now repeat the calculations of the previous section when a dipole $\bf
p$
appears in the center of a cell, with random orientation.  It is convenient
to imagine
the unit sphere centered on the cell partitioned into small solid angle
sections of size $d\Omega$ labelled by $\gamma$, with $\bf p$ only allowed
to take on
values ${\bf p}_{\gamma}$ (pointing to the center of the $\gamma$th
section).
$u_{\alpha, \beta, \gamma}({\bf z},t)=1$ is defined as before,
with the additional implication that ${\bf p}={\bf p}_{\gamma}$.
 The potential at ${\bf x}$
when $t$ lies in the $\beta$th interval is

$$\phi ({\bf x},t)=-G\sum_{\alpha ,\gamma}{1\over{\cal L}^{3}}\int d{\bf z}
u_{\alpha,\beta ,\gamma}({\bf z},t){\bf p}_{\gamma}\cdot
{\bf \nabla}_{z}{1\over |{\bf x}-{\bf z}|}\eqno (5.1)$$

\noindent We note that $<\phi ({\bf x},t)>=0$ since all polarization
directions are equally probable.
Using the same approximations that were made
to obtain Eq. (4.2b), except that ${\cal P}$ need not be small,
the correlation function of the potential is

$$\eqalign{&<\phi ({\bf x},t)\phi ({\bf x}',t')>=\cr
&\qquad\delta (t-t')G^{2}\tilde{\cal P}
\int d{\bf z}\int {d\Omega\over 4\pi}
{\bf p}\cdot{\bf \nabla}_{z}{1\over|{\bf x}-{\bf z}|}
{\bf p}\cdot{\bf \nabla}_{z}{1\over|{\bf x}'-{\bf z}|}\cr
&\qquad\qquad\qquad\qquad
=\delta (t-t'){4\pi\over 3} G^{2}\tilde{\cal P}p^{2}
{1\over |{\bf x}-{\bf x}'|}\cr}\eqno(5.2a,b)$$

	The correlation function of $w_{0}({\bf x},t)=(4\pi
G)^{-1}\nabla^{2}\phi({\bf x},t)$
may now be calculated using Eq. (5.2b):

	$$<w_{0}({\bf x},t)w_{0}({\bf x}',t')>= 3^{-1}\tilde{\cal P}p^{2}\delta
(t-t')
	(-\nabla^{2}) \delta ({\bf x}-{\bf x}') \eqno (5.3)$$

\noindent We see from Eqs. (2.5b) and (2.8) that that these dipole
fluctuations give us the DGGR correlation function with

	$$\gamma '={3\over 16\pi\tilde{\cal P}p^{2}}\eqno(5.3a)$$

\noindent We thus obtain from Eq. (3.5b) an expression for the collapse
rate for a particle of mass m:

 $$\lambda_{m}={m^{2}\over 16\pi^{3\over 2}a\tilde{\cal
P}p^{2}}\eqno(5.4)$$

 	Proceeding exactly as in section 4, we now turn to calculate the rate
 of energy increase of a particle of (smeared) mass m due to the force
 it feels from the fluctuating potential. The correlation function of the
force follows
 from Eq. (5.2b):

 $$\eqalign{<F^{i}(t)F^{j}(t')>&=
\delta (t-t'){4\pi G^{2} m^{2}\tilde{\cal P}p^{2}\over 3(2\pi a^{2})^{3}}
\int d{\bf x}d{\bf x}'
e^{-{1\over 2a^{2}}(x^{2}+x'^{2})}
\partial_{i}\partial '_{j}{1\over |{\bf x}-{\bf x}'|}\cr
&=\delta (t-t')\delta_{ij}
{2\pi^{1\over 2}G^{2}m^{2}\tilde{\cal P}p^{2}\over 9a^{3}}\cr}\eqno
(5.5a,b)$$

\noindent The correlation function in Eq. (5.5b) is that of white noise,
and
we write $F_{i}(t)=KdB_{i}(t)/dt$, where
$K^{2}$ is the constant factor in (5.5b) and $B_{i}(t)$ is Brownian motion.

The momentum is thus $KB_{i}(t)$
and the energy is $E=K^{2}{\bf B}(t)\cdot{\bf B}(t)/2m$, yielding

$${d<E>\over dt}={\pi^{1\over 2}G^{2}m\tilde{\cal P}p^{2}\over 3a^{3}}\eqno
(5.6)$$

\noindent  Equating (5.6) to the CSL rate of energy increase given by (3.4)

 provides our second relationship involving $\lambda_{m}$ and $a$:

$${3\hbar^{2}\over 4ma^{2}}\lambda_{m}={\pi^{1\over 2}G^{2}m\tilde{\cal
P}p^{2}
\over 3a^{3}\hbar}\eqno (5.7)$$

\noindent Eqs. (5.4) and (5.7) may be solved for $\tilde{\cal P}p^{2}$ and
$\lambda_{m}$:

	$$\eqalign{\tilde{\cal P}p^{2}&={3\over 8\pi}\Bigl({\hbar\over G}\Bigr)\cr
	\lambda_{m}&={1\over 6\pi^{1\over 2}}{Gm^{2}\over
a\hbar}\cr}\eqno(5.8a,b)$$

	 The dipole model gives essentially the same expression for $\lambda_{m}
a$ as does
the monopole model (apart from a factor $3^{1\over 2}$, Eq. (5.8b) is the
same as (4.8b)),
and hence essentially the same numerical value (4.9b) for $\lambda_{m}$
when $a$ is taken as $10^{-5}$ cm.  However, Eq. (5.8a) is a requirement
upon
$\tilde{\cal P}p^{2}$ which is independent of  $a$ and $\lambda_{m}$, so
this equation
gives no further information about the collapse parameters. On the other
hand,
if we take the ``natural" values ${\cal T}=$ planck time and
$p=\hbar/c$ (the planck mass times planck length), we find the intriguing
result

	$${{\cal P}\over{\cal L}^{3}}={3\over 8\pi}{1\over(\hbar/\mu
c)^{3}}\eqno(5.9)$$

\noindent which means that, on average, there is always one dipole in a
volume of the order of the
planck volume.

\bigskip

\noindent {\bfg {6. Consistency of Semi-Classical Gravity?}}

\bigskip

	The results we have obtained are based upon presumption of a connection
between collapse
and semi-classical gravity which was suggested by Eq. (1.1).  In sections 4
and 5 gravity
proved fruitful for collapse, and in this section collapse will return the
favor.

	Perhaps the most controversial aspect of semi-classical gravity, as an
approximation
to the ``true" quantum gravity, arises because the expectation value of the
stress tensor
is the gravitational source.  Consider a sphere of matter of radius $R$,
and let the state $|Z>$
describe the sphere with center on the $z$-axis at $z=Z$, and suppose the
state of the
sphere is $(1/\sqrt {2})[|Z>+|-Z>]$.  A probe mass moving along the
$x$-axis will,
according to the standard nonrelativistic quantum theory of gravity, become
entangled with
the state of the sphere, resulting in the statevector $(1/\sqrt
{2})[|Z>|{\rm up}>+|-Z>|{\rm down}>]$,
where $|{\rm up}>$ ($|{\rm down}>$) means that the probe mass is deflected
in the positive (negative)
$z$-direction.  According to semi-classical gravity the probe mass should
be undeflected.
This was actually tested,$^{16}$ with the (not unexpected) result that the
mass is deflected.

	A theoretical objection to semi-classical gravity is that it allows
superluminal communication.$^{17}$  To see this, consider the entangled
state
$(1/\sqrt {2})[|Z>|1>+|-Z>|2>]$, where the states $|1>$ and $|2>$ denote
orthogonal states of
a system which is a large distance from the sphere, but close to a
``sender."
If the sender chooses not to measure the system, the ``receiver," who is
close to the sphere and
uses the probe mass as described above, finds it undeflected.  If, on the
other hand, the sender
chooses to measure the system, thereby finding it to be in state $|1>$ or
$|2>$,
the sphere will immediately be in the state $|Z>$ or $|-Z>$ respectively.
Then the receiver will be able to see this because the probe mass will
now be deflected up or down.

	As we have earlier suggested,$^{2}$ these problems would disappear if the
superposition $(1/\sqrt {2})[|Z>+|-Z>]$ spontaneously collapses to $|Z>$ or
$|-Z>$ before the
probe mass can complete the measurement.  We shall now investigate the
conditions
under which this occurs.

	First, consider the probe particle.  Its uncertainty in position $\Delta
z$
should be of the order of, or less than
$Z$ and its mass should be less than the sphere's mass $\cal M$
in order to obtain an unambiguous deflection indicative of the state of the
sphere.
Thus the probe's velocity uncertainty satisfies

	$$\Delta v_{z}>{\hbar\over {\cal M}Z}\eqno (6.1)$$

	 Now, we are adopting the smearing hypothesis, i.e.,
the gravitational force exerted by each particle is as if the particle's
mass is smeared out over a sphere whose radius
we shall call $a'$: no relation between $a'$ and the GRW parameter $a$
is as yet assumed.   For maximum deflection of the probe we take $Z$ equal
to the sum of the
radii of the {\it effective} spherical mass distributions of the probe and
sphere, so

	$$Z>\max(R,a')\eqno(6.2)$$

	If the probe moves with speed $w$, the time for the measurement to be
performed is
$\approx Z/w$. From Newton's second law we can find the $z$-speed of the
probe, $v_{z}$, if
it is deflected by a sphere at $Z$.   The condition for a good measurement,
capable of detecting
whether the deflection source is one sphere or the superposition, is
$v_{z}>\Delta v_{z}$:

	$${G{\cal M}\over Z^{2}}{Z\over w}>{\hbar\over {\cal M}Z}
	\hbox{\ or\ }\Bigl({{\cal M}\over \mu}\Bigr)^{2}>{w\over c}\eqno (6.3)$$

\noindent We emphasize that, although $Z$ does not directly appear in the
condition (6.3),
it must be restricted as in (6.2) in order that the total mass ${\cal M}$
(and not
a fraction thereof) be the correct mass to appear in (6.3).

	Now, first consider the case $R<a$.
Using Eqs. (3.4) and $\gamma\sim Ga^{2}/\hbar$, $\gamma '\sim G/\hbar$,
we find that the collapse rate for both models is the same,
$\sim G{\cal M}^{2}/a\hbar$.  In order for the outcome of the experiment to

be that the probe is undeflected, the collapse time must be longer than the

time it takes to complete the experiment:

	$${a\hbar\over G{\cal M}^{2}}>{Z\over w}\eqno (6.4)$$

	Combining the inequalities (6.3) and (6.4), we obtain a necessary
condition
for the successful detection of the sphere in a superposed state:

	$$a>Z>a'\eqno(6.5)$$

\noindent (the second inequality in (6.5) comes from (6.2)).
If e.g., $\cal M$ represents some elementary particle and
$a'$ could be e.g., its compton radius, then Eq. (6.5) could easily be
satisfied.
But if the smearing length $a'$ is chosen equal to $a$,
as is mandated by our collapse models, then the inequality  (6.5) cannot
be satisfied.  Thus, in this case, it is impossible to detect
the sphere in a superposition of states by means of the semi-classical
gravitational
force exerted by that superposed state.

	  Lastly, consider the case $R>a$.
Using Eq. (3.4) we find the collapse time for the two models.
The condition that the collapse time be longer than
the time it takes to complete the experiment is

	$$\eqalign{\hbox{GRWP:\qquad}&{\hbar\over G{\cal M}Da^{2}}>{Z\over w}\cr
	\hbox{DGGR:\qquad}&{\hbar R\over G{\cal M}^{2}}>{Z\over
w}\cr}\eqno(6.6a,b)$$

 \noindent Combining the inequalities (6.3) and (6.6), we obtain a
necessary condition
for the successful detection of the sphere in a superposed state:

	$$\eqalign{\hbox{GRWP:\qquad}&{{\cal M}\over Da^{2}}>Z>\max (R,a')\cr
	\hbox{DGGR:\qquad}&R>Z>\max (R,a')\cr}\eqno(6.7a,b)$$

\noindent (the second inequality in Eqs. (6.7) comes from (6.2)).
Eq. (6.7a) can be satisfied for a sufficiently massive object, regardless
of the choice of $a'$, since ${\cal M}\sim R^{3}$.  However, Eq. (6.7b)
cannot be satisfied.

	Thus we conclude, as far as our tentative exploration of the issue is
concerned, that
a CSL collapse model based upon dipole fluctuations may very well allow
semi-classical gravity to
have its cake and eat it too: the metric can be responsive to the
expectation
value of the stress tensor,
yet a nonlocal superposition cannot be detected.  Our investigations
suggest that
it may be worthwhile to look at collapse, aspects of semi-classical
gravity, and
mass-smearing as possible features of quantum gravity.

\bigskip

\noindent {\bfg {Acknowledgments}}

\bigskip

	P. P. would like to thank Henry Stapp for stimulating conversations,
and the Hughes foundation for a supporting grant through Hamilton College.
E. J. S. would like
to thank Chris Dove for discussions.

\bigskip

\noindent {\bfg {References and Remarks}}

\bigskip

1.This prescription is crude because it is
based upon the undefined notion
of measurement by an apparatus to determine which events can occur
and upon the imprecise statement ``after the measurement is completed"
to indicate when the event and its accompanying collapse
take place (note that the Born rule gives, not an {\it absolute}
probability per second
of an event, but rather the {\it conditional} probability of an
event {\it if} one occurs).  See P. Pearle, Amer. Journ of Phys. {\bf 35},
742 (1967);
``True Collapse and False Collapse," to be published in {\it
Quantum-Classical Correspondence}, the
Proceedings of the 4th Drexel Symposium on Quantum Nonintegrability, edited
by Da Hsuan Feng.

2.	P. Pearle, Physical Review A{\bf 39}, 2277 (1989).

3.	G. C. Ghirardi, P. Pearle and A. Rimini, Physical Review A{\bf 42}, 78
(1990).

4.	G. C. Ghirardi and A. Rimini in {\it Sixty-Two Years of Uncertainty},
edited by A. Miller (Plenum, New  York 1990), p. 167;  P. Pearle
in {\it Sixty-Two Years of Uncertainty},
edited by A. Miller (Plenum, New York 1990), p. 193; G. C. Ghirardi and P.
Pearle in
{\it Proceedings of the Philosophy
 of Science Foundation 1990, Volume 2}, edited
 by A. Fine, M. Forbes and L. Wessels (PSA Association, Michigan 1992), p.
19 and p. 35; P. Pearle in
 {\it The Interpretation of Quantum Theory: Where Do We Stand},
edited by L. Accardi (Istituto della Enciclopedia Italiana, Roma 1994), p.
187.

5. L. Ballentine, Phys. Rev. A{\bf 43}, 9 (1991).
The energy creation was realized from the beginning by GRW,$^{6}$ and
indeed
one factor in their choice of parameters was to keep it below
experimental limits.

6.	G. C. Ghirardi, A. Rimini and T. Weber, Physical Review D{\bf 34}, 470
(1986);
Physical Review D{\bf 36}, 3287 (1987); Foundations of Physics {\bf 18}, 1,
(1988);
J. S. Bell in {\it Schrodinger-Centenary celebration of a polymath},
edited by C. W. Kilmister (Cambridge University Press, Cambridge 1987)
and in Speakable and unspeakable in quantum mechanics,
(Cambridge University Press, Cambridge 1987), p. 167.

7.	T. W. B. Kibble, in {\it Quantum Gravity II, A Second Oxford Symposium},
edited by
C. J. Isham, R. Penrose, D. W. Sciama (Clarendon Press, Oxford 1981), p.
63.

8.	L. Diosi, Phys. Rev. A{\bf 40}, 1165 (1989).

9.	G. C. Ghirardi, R. Grassi and A. Rimini, Phys. Rev. A{\bf 42}, 1057
(1990).

10.	This energy gain is compensated by a loss of gravitational
potential energy supplied by the vacuum. The appearance
of e.g., a planck mass monopole means that the vacuum supplies both
the planck mass-energy and the (negative) monopole-particle mutual
gravitational energy.
The energy gain of a particle during its brief period
of acceleration by the monopole comes from a decrease of this mutual
gravitational energy. Thus the subsequent absorption of the monopole by the
vacuum
entails a net loss of gravitational energy of the vacuum.

11.  F. Karolyhazy, Nuovo Cimento {\bf 42}A, 1506 (1966);
 F. Karolyhazy, A Frenkel and B. Lukacs in {\it Physics as Natural
Philosophy},
edited by A. Shimony and H. Feshbach (M.I.T. Press, Cambridge 1982), p.
204;
in {\it Quantum Concepts in Space and Time}, edited by R. Penrose
and C. J. Isham (Clarendon, Oxford 1986), p. 109; A. Frenkel, Found. Phys.
{\bf 20}, 159 (1990).

12. R. Penrose in {\it Quantum Concepts in Space and Time},
edited by R. Penrose and C. J. Isham (Clarendon, Oxford 1986), p. 129; {\it
The Emperor's
New Mind}, (Oxford University Press, Oxford, 1992); {\it
Shadows of the Mind}, (Oxford University Press, Oxford, 1994).

13.	P. Pearle, Physical Review A{\bf 48}, 913 (1993);
"Wavefunction collapse models with nonwhite noise,"  to be published in
{\it Reality and Appearance in Relativistic Quantum Mechanics}, edited by
Rob Clifton
(to be published by Kluwer, 1995).

14.	P. Pearle and E. Squires, Phys. Rev. Lett. {\bf 73}, 1 (1994).

15. B. Collett, P. Pearle, F. Avignone and S. Nussinov, ``Constraint on
Collapse
Models by Limit on Spontaneous X-Ray  Emission in Ge" (preprint, 1994).

16.	D. N. Page and C. D. Geilker, Phys. Rev. Lett {\bf 47}, 979 (1981).

17.	K. Eppley and E. Hannah, Found. Phys. {\bf 7}, 51 (1977).

\end